\documentclass[aps,prl,twocolumn,preprintnumbers]{revtex4}
\usepackage{graphicx}
\usepackage{amssymb}
\usepackage{amsmath}
\newcommand{\be}{\begin{equation}}
\newcommand{\ee}{\end{equation}}
\newcommand{\bea}{\begin{eqnarray}}
\newcommand{\eea}{\end{eqnarray}}
\newcommand{\Eq}[1]{Eq.\,(\ref{#1})}
\newcommand{\Fig}[1]{Fig.\,\ref{#1}}

\newcommand{\GF}{\hat{\mathbf{G}}}

\newcommand{\br}{\mathbf{r}}

\newcommand{\bE}{\mathbf{E}}

\newcommand{\En}{\mathbf{E}_n}

\newcommand{\Em}{\mathbf{E}_{m}}
\newcommand{\Fn}{\mathbf{F}_n}

\newcommand{\heps}{\hat{\boldsymbol{\varepsilon}}}
\newcommand{\hsigma}{\hat{\boldsymbol{\sigma}}}

\newcommand{\omax}{\omega_{\rm max}}

\begin{document}
\title{Resonant-state expansion of dispersive open optical systems}
\author{E.\,A. Muljarov}
\author{W. Langbein}
\affiliation{School of Physics and Astronomy, Cardiff University, Cardiff CF24 3AA,
United Kingdom}
\begin{abstract}
A resonant-state expansion (RSE) for open optical systems with a general frequency dispersion of the relative permittivity, described by a finite number of simple poles, is presented. As in the non-dispersive case, the RSE of dispersive systems converts Maxwell's wave equation into a linear matrix eigenvalue problem in the basis of unperturbed resonant states, in this way numerically exactly determining all relevant eigenmodes of the optical system. This dispersive RSE is verified by application to the analytically solvable system of a sphere in vacuum, with a dispersion of the dielectric constant described by the Drude and Drude-Lorentz models. We calculate the change of the optical modes when converting the sphere material from gold to non-dispersive silica and back to gold, and evaluate the accuracy using the exact solutions.
\end{abstract}
%
%
\date{\today}
\maketitle

Any optical system is characterized by its resonances which are a cornerstone of physics. The concept of resonant states (RSs) is a mathematically rigorous way of treating the resonances. Formally, RSs are the optical eigenmodes of the system, i.e. the eigen-solutions of Maxwell's wave equation, which satisfy the outgoing wave boundary conditions. In open optical systems the RS eigenfrequencies $\omega_n$ are generally complex, which physically reflects the fact that the energy leaks out of the system. The real part ${\rm Re}(\omega_n)$ gives the position of the resonance, while the imaginary part ${\rm Im}(\omega_n)$  gives its half width at half maximum, also determining the quality factor of the resonance as $Q_n=|{\rm Re}(\omega_n)/[2\,{\rm Im}(\omega_n)]|$.

We have recently developed the resonant-state expansion (RSE), a rigorous perturbative method for calculation of RSs, which is treating perturbations of open optical systems of arbitrary strength and shape~\cite{MuljarovEPL10}. We have shown its advantages over established computational methods in electrodynamics, such as finite difference in time domain (FDTD) and finite element method (FEM), in terms of accuracy and efficiency \cite{DoostPRA14}.
Specifically we note that the RSE (i) uses the natural discretization in the frequency domain provided by RSs,
(ii) reduces the solution of Maxwell's wave equation to a linear matrix eigenvalue problem, and
(iii) produces all RSs originating from the basis states in a single calculation, avoiding spurious solutions.
This enables the RSE to determine numerically exactly all the RSs in a frequency range of interest, with the accuracy limited by the basis truncation only.

Other methods use artificial discretization in space and/or in time/frequency domain and the approximation imposed by perfectly matched layers (PMLs) at the system boundaries, both giving rise to issues. FEM, for example, determines RSs one by one, iteratively solving a nonlinear equation with unknown analytics -- it is therefore impractical if not impossible to verify that all RSs within a complex frequency area have been found. In FDTD, RSs can be found by fitting the calculated time evolution by a sum of RSs. Only RSs which have been excited in the simulation are visible, and the fitting procedure does not uniquely determine the number of RSs. Additionally, the spatial discretization and PMLs give rise to spurious solutions.

The reason why the RSE was not available until recently is in the fact that RSs with complex eigenfrequencies have wave functions which are exponentially growing in space away from system, and the proper general normalization of such RSs was not known. The issues with the normalization has been discussed recently, e.g. in Refs.~\cite{SauvanPRL13} and \cite{BaiOE13} where different numerical procedures were suggested. At the same time, the correct normalization corresponding to the spectral representation of the Green's function (GF) of Maxwell's wave equation in terms of RSs is at the heart of the RSE. The correct analytic normalization is contained in our first work on the RSE~\cite{MuljarovEPL10}. This normalization was recently generalized to an arbitrary surface of integration and to optical systems with dispersion which allowed for an exact theory of the Purcell effect \cite{MuljarovARX14}, almost 70 years after its discovery.

So far the RSE has been applied to non-dispersive systems of different dimensionality and geometry~\cite{MuljarovEPL10,DoostPRA12,DoostPRA13,ArmitagePRA14,DoostPRA14}. However, almost all realistic systems, even dielectrics such as glass, have a frequency dispersion of the refractive index.
We have recently found~\cite{DoostARX15}  that the direct substitution of an Ohm's law dispersion into the non-dispersive RSE maintains its linearity. The Ohm's law  dispersion can be a reasonable approximation for some materials whose relative permittivity (RP) is mainly determined by their dc conductivity or when the dispersion can be approximated by a term linear in the light wavelength over the frequency region of interest. However, metals are better described by the Drude model~\cite{JohnsonPRB72}, and a significant improvement is achieved by adding Lorentzian terms~\cite{RakicAO98}, which is further refined by using complex weights (residues) of the frequency poles called critical points (CPs) of the RP~\cite{EtchegoinJCP06,EtchegoinJCP07,VialJPD07}.

\begin{figure}[t]
\includegraphics*[width=\columnwidth]{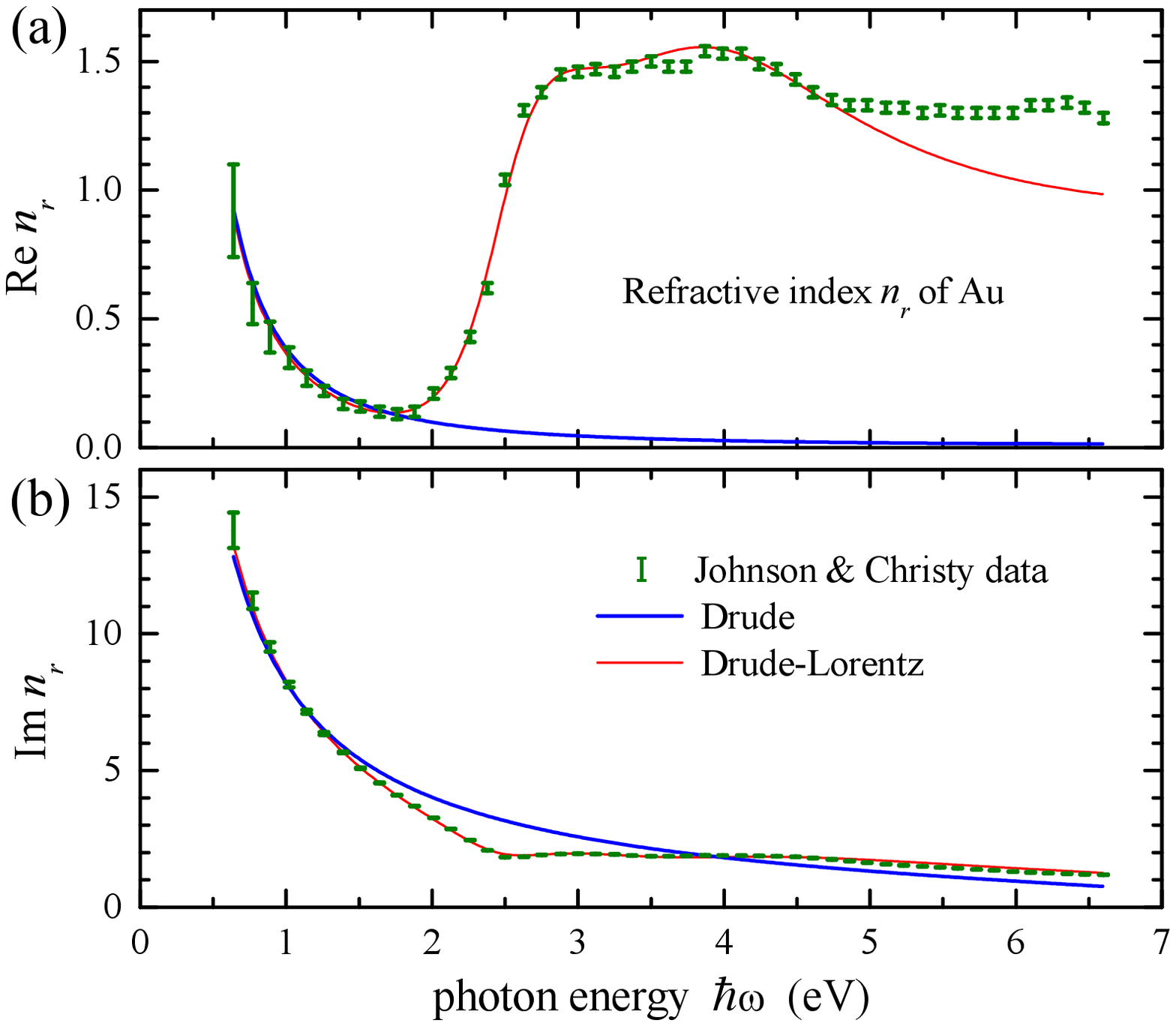}
\caption{(a) real and (b) imaginary parts of the refractive index $n_{r}(\omega)$ of gold, measured by Johnson and Christy~\cite{JohnsonPRB72} (green error bars) and approximated by the Drude model~\cite{SauvanPRL13} (thick blue lines), and by the Drude-Lorentz model with two pairs of CPs~\cite{EtchegoinJCP07} (thin red lines).
}
\label{fig:F1}
\end{figure}

In this Letter we generalize the RSE to the case of a frequency dispersion of the RP with a countable number of poles, suited to describe the RP of any physical material. We verify it on the exactly solvable system of spherical metal and dielectric nano-particles. We start with a dispersive basis of RSs with the wave functions $\En(\br)$ ($\br$ is the position vector) and frequencies $\omega_n$ being the eigen-solutions of Maxwell's wave equation
\be
\label{ME}
\nabla\times\nabla\times\En(\br)=\frac{\omega_n^2}{c^2}\,\heps(\br,\omega_n)\En(\br)
\ee
and the electric fields $\En(\br)$ satisfying the outgoing wave boundary conditions~\cite{DoostPRA13}. The dispersive RP tensor $\heps(\br,\omega)$ of an unperturbed open optical system is taken in the form of a function in the complex frequency plane expressed as
\be
\heps(\br,\omega)=\heps_\infty(\br)+\sum_j\frac{i\hsigma_j(\br)}{\omega-\Omega_j}\,,
\label{eps}
\ee
where $\heps_\infty(\br)$ is the high-frequency value of the RP and $\Omega_j$ are the  resonance frequencies (poles) of the RP determining the dispersion, with the weight tensors $\hsigma_j(\br)$ corresponding to generalized conductivities of the medium at these resonances. The Lorentz reciprocity theorem requires that all tensors in \Eq{eps} are symmetric, and the causality principle requires that $\heps^\ast(\br,\omega)=\heps(\br,-\omega^\ast)$ ~\cite{LandauLifshitzV8Book84}. Therefore, for a physically relevant dispersion, each pole of the RP with a positive real part of $\Omega_j$ has a partner at $\Omega_{-j}=-\Omega^\ast_j$ with $\hsigma^\ast_{-j}=\hsigma_j$, while poles with zero real part of $\Omega_j$ have real $\hsigma_j$. The Ohm's law dispersion of the RP corresponds to the sum in \Eq{eps} replaced by a single term  with $\Omega_0=0$ and $\hsigma_0(\br)$ being the dc conductivity tensor.  The Drude model of metals consists of two poles with $\Omega_0=0$, $\Omega_1=-i\gamma$, and $\hsigma_1(\br)=-\hsigma_0(\br)$. The Drude-Lorentz model introduces additional  poles at $\omega=\Omega_j$ with $j=\pm2,\pm3,\dots$ and  complex conductivities $\hsigma_j$. \Fig{fig:F1} provides an example of the Drude and Drude-Lorentz models approximating the measured complex refractive index $n_{r}(\omega)=\sqrt{\epsilon(\omega)}$ of gold~\cite{JohnsonPRB72}, with the parameters taken from Refs.~\cite{SauvanPRL13} and \cite{EtchegoinJCP07}, respectively,
and used in the following for illustration of the dispersive RSE. In particular, we used for the Drude model $\hbar\Omega_1=-92.8\,i$\,meV, $\hbar\sigma_1=-744$\,eV, and $\varepsilon_\infty=1$~\cite{SauvanPRL13}. For the Drude-Lorentz model we used $\hbar\Omega_1=-85.6\,i$\,meV, $\hbar\sigma_1=-882$\,eV, $\hbar\Omega_2=(2.64-0.65\,i)$\,eV, $\hbar\sigma_2=3.35\,e^{i\pi/4}$\,eV, $\hbar\Omega_3=(3.82-1.17\,i)$\,eV, $\hbar\sigma_3=4.20\,e^{i\pi/4}$\,eV, and $\varepsilon_\infty=1.54$~\cite{EtchegoinJCP07}.

The GF of Maxwell's wave equation has an infinite countable number of simple poles in the complex frequency plane, and therefore has the spectral representation
\be
\GF_\omega(\br,\br')=c^2\sum _n\frac{\En(\br)\otimes\En(\br')}{2\omega_n(\omega-\omega_n)}\,,
\label{ML}
\ee
where the sum is taken over all RSs, and $\otimes$ denotes the dyadic product of vectors. The spectral representation \Eq{ML} requires that the RSs are normalized according to ~\cite{MuljarovARX14}
\bea
\!\!\!\!\!\!\!\!\!\!1+\delta_{0,\omega_n}\!\!\!&=&\!\!\!\int_{V}\En (\br)\cdot\left.\frac{\partial\bigl(\omega^2\heps(\br,\omega)\bigr)}{\partial(\omega^2)}\right|_{\omega_n}\!\!\En(\br)\,d{\bf r}
\nonumber
\\
&&\!\!\!+\frac{c^2}{2\omega^2_n}\oint_{S_V} \left(\En\cdot\frac{\partial\Fn}{\partial
s}-\Fn \cdot \frac{\partial \En}{\partial
s}\right) dS\,,
 \label{norm}
\eea
where $\Fn=(\br\cdot\nabla)\En$, $V$ is an arbitrary simply connected volume with a boundary surface $S_V$ enclosing the inhomogeneity of the system, and the derivative
$\partial/\partial s$ is taken along the outer surface normal. Note that for static modes ($\omega_n=0$), the volume of integration can be extended to the entire space and the surface term vanishes, since the electric field in such modes decay or just vanishes outside the system. All other modes instead have an electric field exponentially growing outside the system, so that the normalization has to be evaluated for a finite $V$. Substituting \Eq{ML} into Maxwell's wave equation for the GF and using \Eq{ME} we obtain
\be
\sum_n\frac{\omega^2\heps(\br;\omega)-\omega_n^2\heps(\br;\omega_n)}{2\omega_n(\omega-\omega_n)}\,\En(\br)\otimes\En(\br')=
\hat{\mathbf 1}\delta(\br-\br')
\label{Closure}
\ee
which has to be satisfied for any $\omega$. For the dispersion of the RP in the form of \Eq{eps}, we find with the help of the algebraic identity
\bea
&&\frac{1}{\omega_n(\omega-\omega_n)} \left(\frac{\omega^2}{\omega-\Omega_j}-\frac{\omega_n^2}{\omega_n-\Omega_j}\right)\nonumber\\
&&=\frac{1}{\omega_n-\Omega_j}+\frac{\omega}{\omega-\Omega_j}\left(\frac{1}{\omega_n}-\frac{1}{\omega_n-\Omega_j}\right)
\label{algebra1}
\eea
that \Eq{Closure} splits into the following {\it closure relation}
\be
\frac{1}{2}\sum_n\heps(\br,\omega_n)\,\En(\br)\otimes\En(\br')=
\hat{\mathbf 1}\delta(\br-\br')
\label{Closure2}
\ee
and {\it sum rules}
\be
\sum_n\,\frac{\En(\br)\otimes\En(\br')}{\omega_n-\Omega_j}=0\,.
\label{sum}
\ee
Now, again using the algebraic identity
\be
\frac{1}{\omega_n(\omega-\omega_n)}-\frac{1}{\omega\omega_n}+\frac{\Omega_j}{\omega^2(\omega_n-\Omega_j)}=\frac{W_n^j(\omega)}{\omega(\omega-\omega_n)}\,,
\label{algebra2}
\ee
where
\be
W_n^j(\omega)=\frac{\omega_n}{\omega}\,\frac{\omega-\Omega_j}{\omega_n-\Omega_j}\,,
\ee
and combining \Eq{ML} with sum rules \Eq{sum} for $\Omega_0=0$ and  $\Omega_j\neq0$ according to the terms in \Eq{algebra2},
we find an additional spectral representation $\GF^j_\omega$ of the GF for each pole in the RP:
\be
\GF^j_\omega(\br,\br')=c^2\sum _n W_n^j(\omega)\frac{\En(\br)\otimes\En(\br')}{2\omega(\omega-\omega_n)}\,.
\label{GF1}
\ee
The Ohm's law dispersion introduces a $\omega=0$ pole in the RP which leads to the sum rule \Eq{sum} corresponding to $\Omega_0=0$. This sum rule results in the representation $\GF^0_\omega(\br,\br')$ of the GF given by \Eq{GF1} with $W_n^0(\omega)=1$. Note however, that the $\omega=0$ pole is implicitly present already in the non-dispersive system owing to the longitudinal $\omega_n=0$ modes~\cite{DoostPRA14}. As a result, the sum rule \Eq{sum} with $\Omega_0=0$ holds also without dispersion~\cite{DoostPRA13,ArmitagePRA14}, due to the constant term in the RP, such as $\heps_\infty(\br)$ in \Eq{eps}. This explains why Ohm's law does not need any significant reformulation of the RSE compared to the non-dispersive case and does not require an extension of the basis of RSs~\cite{DoostARX15}.

We now solve a perturbed Maxwell's wave equation equivalent to \Eq{ME}, with the unperturbed RP $\heps(\br,\omega)$ replaced by a perturbed one, $\heps(\br,\omega)+\Delta\heps(\br,\omega)$, with the perturbation $\Delta\heps(\br,\omega)$ of the form of \Eq{eps} described by the  perturbations $\Delta\heps_\infty(\br)$ of $\heps_\infty(\br)$ and $\Delta\hsigma_j(\br)$ of $\hsigma_j(\br)$ inside the unperturbed system. We find the electric field $\bE(\br)$ and the eigenfrequency $\omega$ of a perturbed RS by using the unperturbed GF in different representations \Eq{GF1} for the corresponding terms of the RP, yielding
\bea
\bE(\br)&=&-\frac{\omega^2}{c^2} \int \GF_\omega(\br,\br') \Delta\heps(\br',\omega) \bE(\br') d\br'
\nonumber\\
&=&-\frac{\omega^2}{c^2} \int \biggl[\GF^0_\omega(\br,\br') \Delta\heps_\infty(\br',\omega) \nonumber\\
&&+ \sum_j \GF^j_\omega(\br,\br') \frac{i\Delta\hsigma_j(\br',\omega)}{\omega-\Omega_j}\biggr] \bE(\br') d\br'\,.
\label{E2}
\eea
This integral equation is then converted to a matrix equation by expanding the perturbed RS into the basis of unperturbed ones,
\be
\bE(\br) = \sum_n c_{n}\En(\br)\,,
\ee
by using expansions \Eq{GF1} of the GF, and by equating the coefficients at different basis functions $\En(\br)$. The result is the eigenvalue equation
\be
(\omega_n-\omega)\sum _{m} \Bigl[2\delta_{nm}+V_{nm}(\infty)\Bigr]c_{m}=\omega_n \sum _{m} V_{nm}(\omega_n)c_{m}
\label{RSE1}
\ee
which is linear in $\omega$, with the perturbation matrix
\be
V_{nm}(\omega)=\int \En(\br)\cdot\Delta\heps(\br,\omega)\Em(\br)\,d \br\,.
\label{Vnm}
\ee
This is the {\it linear dispersive RSE}.  The perturbation matrix $V_{nm}(\omega)$ represents the change $\Delta\heps(\br,\omega)$ of the RP for any physical dispersion described by \Eq{eps}. In the absence of dispersion, $V_{nm}(\omega)=V_{nm}(\infty)$, and \Eq{RSE1} simplifies to\break $2(\omega_n-\omega)c_{n}= \omega\sum _{m} V_{nm}c_{m}$ which is the eigenvalue equation of the non-dispersive RSE~\cite{MuljarovEPL10,DoostPRA14}.

The linear dispersive RSE is suited for both dispersive and non-dispersive unperturbed systems with perturbations which do or do not increase the number of poles in the RP. When increasing the number of non-zero poles of the RP, the number of poles of the GF is increased too~\cite{BroerJPA09}, and the size of the RSE basis is extended by an additional countable infinite number of RSs for each non-zero pole of the RP, with the RS eigenfrequencies asymptotically approaching this pole. Poles of the RP with finite weight in the perturbed system but zero weight in the unperturbed system are included in the basis by taking the limit of the pole weight tending to zero. In this limit, the pole-related RSs have frequencies converging to the pole but refractive indices taking separate discrete values, as detailed below.

\begin{figure}[t]
\includegraphics*[width=\columnwidth]{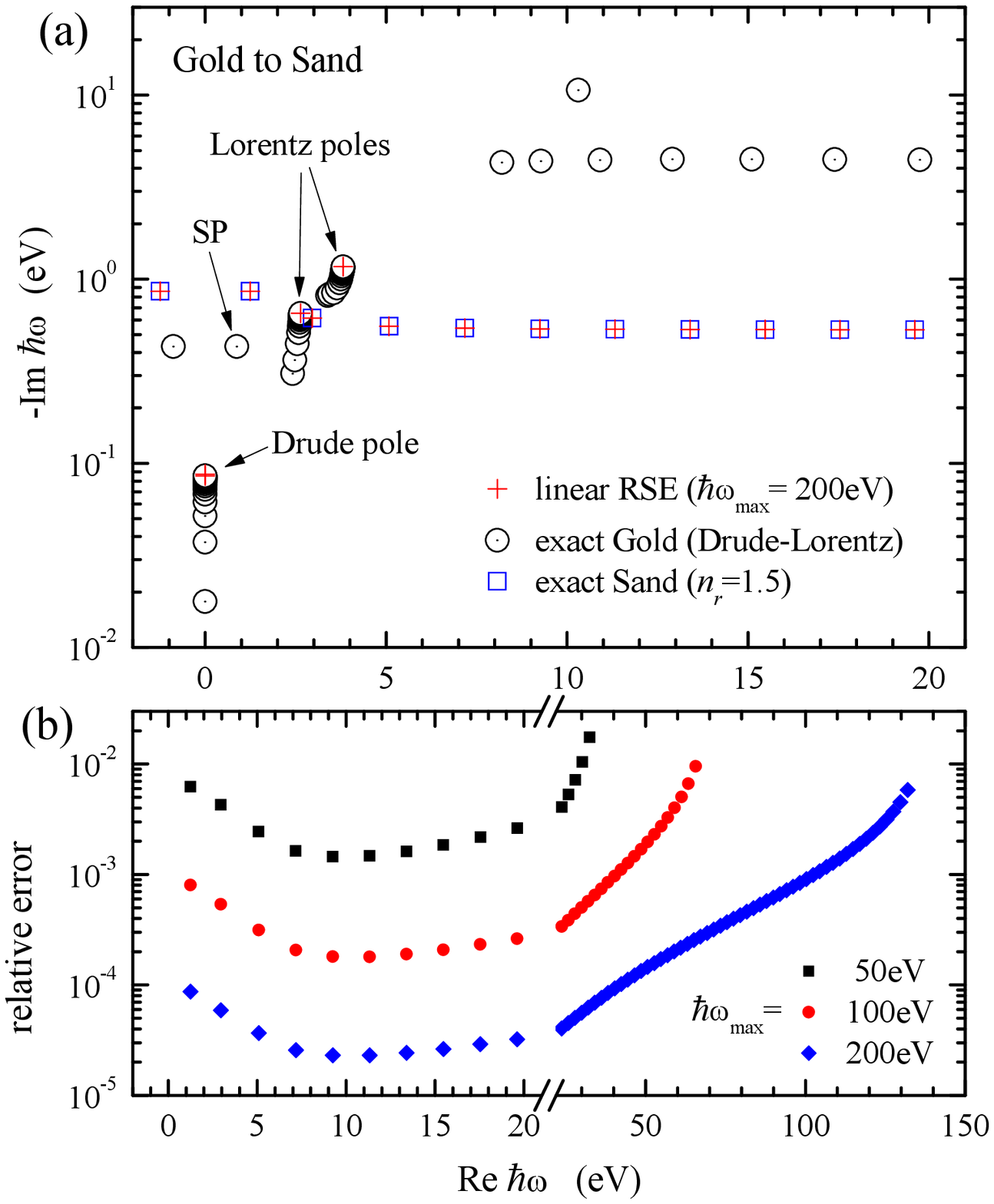}
\caption{Results of the dispersive RSE converting gold into sand. (a) RS energies $\hbar\omega_n$ of the unperturbed system (gold sphere in vacuum, black circles with dots) and the perturbed system (sand sphere in vacuum) for $l=1$ TM modes and the sphere radius of $R=200$\,nm. The perturbed energies are calculated exactly (blue squares) and using the linear RSE \Eq{RSE1} (red crosses) for $\hbar\omax=200$\,eV. (b) Relative difference between the RSE and exact eigenenergies, for different values of $\omax$ as given. The RP of gold was modelled with the Drude-Lorentz model with two pairs of CPs and parameters taken from Ref.~\cite{EtchegoinJCP07}, while $n_{r}=1.5$ was used for sand.
}
\label{fig:F2}
\end{figure}

\begin{figure}[t]
\includegraphics*[width=\columnwidth]{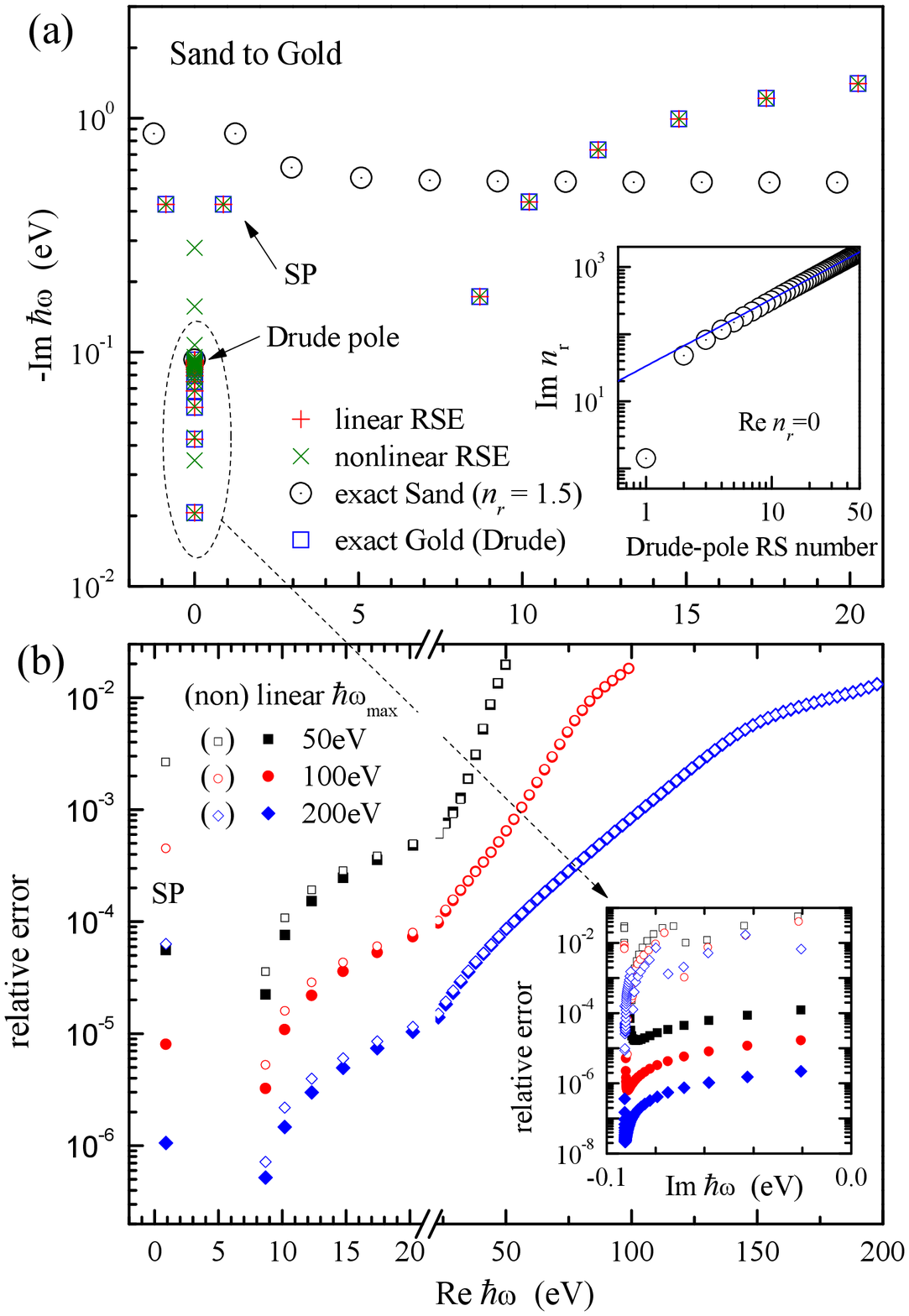}
\caption{Results of the dispersive RSE converting sand into gold. (a) RS energies $\hbar\omega_n$ of the unperturbed system (sand sphere with $n_{r}=1.5$ in vacuum, black circles with dots) and perturbed system (gold sphere in vacuum) for $l=1$ TM modes and $R=200$\,nm. The RP of gold was modelled with the Drude model with parameters taken from Ref.~\cite{SauvanPRL13}. The perturbed energies are calculated exactly (blue open squares), and using the linear RSE \Eq{RSE1} (red crosses) or the nonlinear RSE \Eq{RSE2} (green crosses). The inset shows the refractive index of degenerate Drude-pole modes in the unperturbed basis sorted in ascending order. The line shows a proportionality between number and index. (b) Relative error of the RSE energies of the perturbed RSs, for both the linear and nonlinear RSE, for different values of $\omax$ as given. The inset shows the relative errors of RSs close to the Drude pole which have purely imaginary energies.
}
\label{fig:F3}
\end{figure}

To illustrate the method and evaluate its convergence, we show in Figs.~\ref{fig:F2} and \ref{fig:F3} the transverse magnetic (TM) eigenmodes with $l=1$ ($l$ is the orbital number) of spheres made of a dispersive material (gold) and a non-dispersive material (sand, $n_{r}=1.5$) in vacuum, and perturbations which transform gold to sand in Fig.~\ref{fig:F2} and sand to gold in Fig.~\ref{fig:F3}. The eigenmodes of the sand and gold spheres in vacuum were taken in the analytic form~\cite{DoostPRA14} and normalized according to \Eq{norm}, see Ref.~\cite{MuljarovARX14} for explicit analytic expressions. The radius of the sphere $R=200$\,nm is chosen such that both Drude and Drude-Lorentz approximations of the gold dispersion shown in Fig.~\ref{fig:F1} are valid for the frequency of the fundamental surface plasmon (SP) mode shown in Figs.~\ref{fig:F2} and \ref{fig:F3} by arrows.

We select a finite number $N$ of RSs for the RSE basis, including only RSs satisfying the condition $|n_r(\omega_n)\omega_n|<\omax$. This excludes RSs having a wavevector in the medium above  $\omax/c$, which is the case of large $\omega_n$ or large $|n_r(\omega_n)|$ with $\omega_n$ close to the poles of the dispersion. This basis selection can be optimized in the future. The RSE results for the perturbed eigenmodes are compared with the analytic solutions, and the relative errors are shown in Figs.\,\ref{fig:F2}(b) and \ref{fig:F3}(b) for different $\omax$, demonstrating a high accuracy given the strong perturbation. For the present geometry, $N$ is approximately proportional to $\omax$, with $N=456$ for $\hbar\omax=200\,$eV. The observed $1/N^3$ convergence to the exact solution is comparable to that of the non-dispersive RSE~\cite{MuljarovEPL10,DoostPRA14}.

Going from gold to sand [Fig.~\ref{fig:F2}(a)] the RSE reproduces the RSs of the non-dispersive sand sphere, and additionally produces a number of quasi-degenerate RSs at the Drude and Lorentz poles. These RSs are present in the system since in the linear RSE the same poles of the dispersion are present before and after perturbation. Poles which have zero weight in a system still lead to an infinite series of RSs, with frequencies at the pole position, but corresponding to different refractive indices, as exemplified in the inset of Fig.~\ref{fig:F3}(a). For the sphere geometry, they can be calculated analytically by taking the limit of the pole weight to zero in the secular equation. A perturbation which creates a finite weight of the pole lifts the degeneracy of these RSs as exemplified in Fig.~\ref{fig:F3}.

We now compare this result with an alternative dispersive RSE approach which uses a non-dispersive system as basis and creates the additional RSs due to the poles of the dispersion via the nonlinearity of the resulting generalized eigenvalue problem. Assuming that the unperturbed $\heps(\br)$ has no dispersion, the only valid sum rule in \Eq{sum} is the one with $\Omega_j=0$ which provides only one alternative GF representation $\GF^0_\omega(\br,\br')$ with $W^{0}_n(\omega)=1$ in \Eq{GF1}. Replacing $\GF^j_\omega$ by $\GF^0_\omega$ in \Eq{E2} results in a {\it nonlinear dispersive RSE}
\be
2(\omega_n-\omega)c_n= \omega\sum _{m} V_{nm}(\omega)c_m\,,
\label{RSE2}
\ee
which is a direct generalization of the original, non-dispersive RSE~\cite{MuljarovEPL10,DoostPRA14}. For a finite number of poles in the RP, \Eq{RSE2} can be written as a polynomial matrix equation. The order $M$ of the polynomial is given by the number of non-zero poles in \Eq{eps}, so for example, $M=1$ for the Ohm's law model (linear matrix problem), $M=2$ for the Drude model (quadratic problem), and $M=6$ for the Drude-Lorentz model with 2 pairs of CPs. For any finite $M>1$, such a polynomial eigenvalue problem can be solved by linearization~\cite{HighamSJMAA06}, extending the basis of unperturbed RSs by a factor of $M$.

We illustrate this alternative method for the Drude dispersion of the perturbed system, for which  \Eq{RSE2} is a quadratic matrix problem. For the same basis cut-off $\omax$ as used for the linear dispersive RSE, the energies of the Fabry-P\'erot RSs are reproduced with a similar accuracy, see Fig.~\ref{fig:F3}(b). However, the SP mode has about 2 orders of magnitude larger error and modes around the Drude pole are also having orders of magnitude larger error as shown in the inset of \Fig{fig:F3}(b). This can be understood considering that in the nonlinear RSE the basis does not contain the pole RSs, and is therefore less suited to describe the RSs close to the RP poles.

In conclusion, the presented generalization of the RSE to materials with an arbitrary frequency dispersion of the relative permittivity, described by a finite number of simple poles, is extending the applicability of the RSE to all relevant open optical systems, paving the way to its widespread use in electromagnetic simulation.

\acknowledgments This work was supported by the Cardiff University EPSRC Impact Acceleration Account EP/K503988/1, and the S\^er Cymru National Research Network in Advanced Engineering and Materials. The authors acknowledge discussions with M.\,D. Doost and H. Sehmi.


\begin{thebibliography}{10}

\bibitem{MuljarovEPL10}
E.~A. Muljarov, W. Langbein, and R. Zimmermann, Europhys. Lett. {\bf 92},
  50010  (2010).

\bibitem{DoostPRA14}
M.~B. Doost, W. Langbein, and E.~A. Muljarov, Phys. Rev. A {\bf 90},  013834
  (2014).

\bibitem{SauvanPRL13}
C. Sauvan, J.~P. Hugonin, I.~S. Maksymov, and P. Lalanne, Phys. Rev. Lett. {\bf
  110},  237401  (2013).

\bibitem{BaiOE13}
Q. Bai {\it et~al.}, Opt. Express {\bf 21},  27371  (2013).

\bibitem{MuljarovARX14}
E. Muljarov, M. Doost, and W. Langbein, arXiv  1409.6877  (2014).

\bibitem{DoostPRA12}
M.~B. Doost, W. Langbein, and E.~A. Muljarov, Phys. Rev. A {\bf 85},  023835
  (2012).

\bibitem{DoostPRA13}
M.~B. Doost, W. Langbein, and E.~A. Muljarov, Phys. Rev. A {\bf 87},  043827
  (2013).

\bibitem{ArmitagePRA14}
L.~J. Armitage, M.~B. Doost, W. Langbein, and E.~A. Muljarov, Phys. Rev. A {\bf
  89},  053832  (2014).

\bibitem{DoostARX15}
M. Doost, W. Langbein, and E. Muljarov, arXiv  1508.03851v1  (2015).

\bibitem{JohnsonPRB72}
P.~B. Johnson and R.~W. Christy, Phys. Rev. B {\bf 6},  4370  (1972).

\bibitem{RakicAO98}
A.~D. Raki\'{c}, A.~B. Djuri$\check{\rm s}$i\'{c}, J.~M. Elazar, and M.~L.
  Majewski, Appl. Opt. {\bf 37},  5271  (1998).

\bibitem{EtchegoinJCP06}
P.~G. Etchegoin, E.~C. Le~Ru, and M. Meyer, J. Chem. Phys. {\bf 125},  164705
  (2006).

\bibitem{EtchegoinJCP07}
P.~G. Etchegoin, E.~C. Le~Ru, and M. Meyer, J. Chem. Phys. {\bf 127},  189901
  (2007).

\bibitem{VialJPD07}
A. Vial and T. Laroche, J. Phys. D: Appl. Phys. {\bf 40},  7152  (2007).

\bibitem{LandauLifshitzV8Book84}
L.~D. Landau, L.~P. Pitaevskii, and E. Lifshitz, {\em Electrodynamics of
  Continuous Media}, Vol.~8 of {\em Course of Theoretical Physics}, 2nd ed.
  (Elsevier Butterworth Heinemann, Oxford, 1984), iSBN-10: 0750626348.

\bibitem{BroerJPA09}
W. Broer and B.~J. Hoenders, J. Phys. A: Math. Theor. {\bf 42},  245207
  (2009).

\bibitem{HighamSJMAA06}
J. Higham, D.~S. Mackey, N. Mackey, and F. Tisseur, SIAM J. Matrix Anal. Appl.
  {\bf 29},  143  (2006).

\end{thebibliography}
\end{document}